\documentclass[runningheads,a4paper]{llncs}

\usepackage{amssymb}
\usepackage{graphicx}

\usepackage{url}
\urldef{\mailsa}\path|{daniel.stein@nyu.edu, newman@cims.nyu.edu}|    
\newcommand{\keywords}[1]{\par\addvspace\baselineskip
\noindent\keywordname\enspace\ignorespaces#1}

\begin{document}

\mainmatter  

\title{Nature vs.~Nurture in Complex and Not-So-Complex Systems}

\titlerunning{Nature vs.~Nurture}

%
%
\author{D.L.~Stein
\thanks{Also at Department of Physics, New York University, NY, NY USA 10003}
\and C.M.~Newman
\thanks{Also at Department of Mathematics, University of California, Irvine, CA USA 92697}}
\authorrunning{Stein and Newman}

\institute{Courant Institute of Mathematical Sciences\\
New York University, New York, NY USA 10011\\}

%
%

\toctitle{Lecture Notes in Computer Science}
\tocauthor{Authors' Instructions}
\maketitle

\begin{abstract}
  Understanding the dynamical behavior of many-particle systems both
  in and out of equilibrium is a central issue in both statistical
  mechanics and complex systems theory. One question involves ``nature
  vs.~nurture'': given a system with a random initial state evolving
  through a well-defined stochastic dynamics, how much of the
  information contained in the state at future times depends on the
  initial condition (``nature'') and how much on the dynamical
  realization (``nurture'')? We discuss this question and present both
  old and new results for low-dimensional Ising spin systems.

\keywords{heritability, persistence, aging, damage spreading, Ising spin dynamics}
\end{abstract}

\section{Introduction}

The nonequilibrium dynamics of both thermodynamic and complex systems
(the intersection of these two sets is nonempty) remains an area of
intensive research, and a host of open problems remains. The most
extreme case of nonequilibrium dynamics occurs after a {\it deep
  quench\/}, in which a system in equilibrium at a very high
temperature is instantaneously cooled to a very low temperature, after
which it evolves according to a well-defined dynamics corresponding to
that low temperature.  The extreme case of a deep quench is the
instantaneous cooling of a system from infinite to zero
temperature. The subsequent zero-temperature dynamics consists of the
system's running ``downhill'' in energy (or uphill in survival
probability, if one is dealing with a biological system) to some local
or global minimum (or maximum).

Determining the state of such a system at long times, given both the
initial state and the subsequent dynamics, is a difficult --- and
generally unsolved --- problem, even for relatively simple systems. In
this paper, we will review progress on this question for Ising spin
systems, both homogeneous and disordered, in one and two
dimensions. We will see that even in $2D$ the problem is far from
simple, with open questions remaining even for --- in fact, especially
for --- the uniform ferromagnet. However, recent progress has been
made, and the insights gained may be useful in understanding dynamical
properties of more interesting --- and possibly complex --- systems.

\section{Types of Long-Time Behavior}

For concreteness we consider an Ising~spin system on the infinite
lattice ${\bf Z}^d$; that is, at each site $x\in{\bf Z}^d$ we assign a
binary variable $\sigma_x=\pm 1$. We restrict our attention to models
in which the spin-spin couplings are nearest-neighbor. The most basic
question one might ask is whether, after a deep quench, the dynamics
eventually settles down to a fixed state, or whether some or all spins
continue to flip forever. 

The notion of equilibration of an infinite system after a finite time
contains some subtleties, which we will address in the next
section. But without addressing these subtleties, we can pose the
question in a precise way: does the spin configuration have a limit as
$t\to\infty$?  Equivalently, for every $x$, does $\sigma_x$ flip
infinitely often or only finitely many times? (Note that even for
those systems in which the latter is true, it will generally {\it
  not\/} be the case that there exists some finite time $T_0$ after
which every spin has stopped flipping. This is discussed further in
the next section.) From this perspective, it is useful to distinguish
among three classes of dynamical system: a system is type~${\cal F}$
if every spin flips only finitely many times; type~${\cal I}$ if every
spin flips infinitely often; and type~${\cal M}$ (for ``mixed'') if
some spins flip infinitely often and others do not~\cite{NS00}. The
overall spin configuration has a limit only of the system is
type~${\cal F}$.

Determining which class a system belongs to is generally a nontrivial
problem; in fact, the answer remains unknown even for uniform
ferromagnets (and antiferromagnets) in ${\bf Z}^d$ for $d\ge 3$. There
does exist some numerical work, however, suggesting that these might
be type~${\cal I}$ for $d=3$ and $4$, but type~${\cal F}$ --- or
possibly ${\cal M}$ --- for $d\ge 5$~\cite{Stauffer94}.

However, some progress has been made. There exist proofs that uniform
ferromagnets or antiferromagnets in one dimension (on ${\bf Z}$) and
in two dimensions (on ${\bf Z}^2$) are type~${\cal I}$. Moreover, in
{\it any\/} dimension on a lattice where each site has an odd number
of nearest neighbors, they are type~${\cal F}$~\cite{NNS00}. Work has
also been done on two-dimensional ``slabs'': that is, systems that are
infinite in two dimensions but consist of a finite number of layers in
the third. Here the system can be either type~${\cal F}$ or~${\cal
  M}$, depending both on the number of layers and on the boundary
conditions (free or periodic) in the third (finite) direction. For
details, see~\cite{DKNS13}.

It was further proved in~\cite{NNS00} that all models with continuous
disorder, in which the spin-spin couplings are chosen from a common
distribution with finite mean, belong to class~${\cal F}$ in all
dimensions and on all types of lattice. These include ordinary
Edwards-Anderson spin glasses~\cite{EA75} and random ferromagnets.  We
ignore here systems with continuous disorder in which the distribution
has infinite mean, and refer the interested reader
to~\cite{NS00,NN99}.

Another class of systems comprises the so-called $\pm J$~spin glass
models, where each coupling independently takes on the value $+J$ or
$-J$ with equal probability. It was shown that in one dimension these
are type~${\cal I}$, and in two dimensions (again on ${\bf Z}^2$)
type~${\cal M}$~\cite{GNS00}.  And once again, on any lattice
regardless of dimension where each site has an odd number of
neighbors, they are type ${\cal F}$.

Results exist also for systems with more exotic coupling
distributions; we refer the interested reader to~\cite{NS00}.  We now
turn to the next question, which is our main interest here: what can
be learned about the state of a system at a finite time $t$ after a
deep quench. The answer, not surprisingly, depends on which class the
system belongs to, but as we shall see, in most cases one is forced to
undertake numerical simulations to gain insight.

\section{Local Equilibration, Local Non-Equilibration, and Chaotic
  Size Dependence}

How might one think about equilibration in an infinite system, even
one of type-${\cal F}$, given that at any finite time some spins still
not have reached their final state? It was proposed in~\cite{NS99}
that this problem could be understood in the sense of {\it local
  equilibration\/}: choose a region of fixed size surrounding the
origin, and ask whether, after a finite time, domain walls cease to
sweep across the region, overturning the spins within. This timescale
$\tau(L)$ is expected to increase without bound as $L$ goes to
infinity (and in general will also depend on the choice of initial
condition, dynamics, lattice type and dimensionality, and possibly
other factors); but the idea is that as long as $\tau(L)<\infty$ for
{\it any\/} $L<\infty$, no matter how large, then we can say that the
system undergoes local equilibration. Any system of type-${\cal F}$
obviously undergoes local equilibration. Types~${\cal I}$ and ${\cal
  M}$ do not, and we say that these systems experience {\it local
  nonequilibration\/} (LNE)~\cite{NS99}.

LNE can be of two types. Even though the configuration in a given
finite region never settles down, one can still ask whether, if one
averages over all dynamical realizations, the {\it dynamically
  averaged\/} configuration settle down to a limit at large times. Or
does even this averaged configuration not settle down?

The first possibility (a limit of the dynamically averaged
configuration) can be thought of as ``weak LNE'', while the second (no
limit) is referred to as {\it chaotic time dependence\/}
(CTD)~\cite{NS99}. As shown in~\cite{NS99}, weak~LNE implies a
complete lack of predictability (nurture ``wins'' --- after some time,
the dynamics wipes out information about the initial state), while CTD
implies that some amount (which can be quantified) of predictability
remains (nature wins).

So a study of nature vs.~nurture provides a great deal of information
on a number of central dynamical issues concerning classes of
dynamical systems. We now review both older and more recent results
for different Ising-like spin systems, both homogeneous and
disordered.
  
\section{Nature vs.~Nurture in $1D$ Random Ferromagnets and Spin
  Glasses}

Because type-${\cal F}$ models always equilibrate locally, one can
simply compare the final state of a spin with its initial state over
many dynamical trials to determine whether initial information has
been fully retained, partially retained, or completely lost. This can
be quantified by introducing~\cite{NNS00} a type of dynamical order
parameter, denoted $q_D$, that in some ways serves an analogue of the
(equilibrium) Edwards-Anderson order parameter~$q_{EA}$~\cite{EA75}.

Let $\sigma^t$ denote the (infinite-volume) spin configuration at time
$t$ given a specific initial configuration $\sigma^0$ and dynamical
realization $\omega$ (for notational convenience, the dependence of
$\sigma^t$ on $\sigma^0$ and $\omega$ is suppressed). We want to
study, for fixed $\sigma^0$ (and, if the model is disordered, fixed
coupling realization~${\cal J}$), this quantity averaged over all
dynamical realizations up to time $t$; denote such an average by by
$\langle \cdot \rangle_t$. One then needs to study the resulting
quantity averaged over all initial configurations and coupling
realizations. Denoting the latter averages (with respect to the joint
distribution $P_{{\cal J},\sigma^0} = P_{\cal J} \times P_{\sigma^0}$)
by ${\bf E}_{{\cal J},\sigma^0}$, we define $q_D = \lim_{t \to
  \infty}q^t$ (providing the limit exists), where
\begin{equation}
\label{eq:qt}
q^t = \lim_{L \to \infty} |\Lambda_L|^{-1}\sum_{x \in \Lambda_L}
(\langle \sigma_x \rangle_t)^2 = {\bf E}_{{\cal J},\sigma^0}
(\langle \sigma_x \rangle_t^{\,2})
\end{equation}
and $\Lambda_L$ is a $d$-dimensional cube of side $L$ centered at the
origin.  The equivalence of the two formulas for $q^t$ follows from
translation-ergodicity~\cite{NNS00}.

The order parameter $q_D$ measures the extent to which $\sigma^\infty$
is determined by $\sigma^0$ rather than by $\omega$.  It was proved
in~\cite{NNS00} that for the $1D$ random ferromagnet and/or spin glass
with continuous disorder, $q_D=1/2$.  What this means is that, for
a.e.~${\cal J}$ and $\sigma^0$, precisely half of the $x$'s in ${\bf
  Z}$ have $\sigma_x^\infty$ completely determined by $\sigma^0$ with
the other $\sigma_x^\infty$'s completely undetermined by $\sigma^0$.

We turn now to the more difficult case of type-${\cal I}$ systems.  It
seems somewhat counterintuitive that models with continuous disorder,
in particular random ferromagnets and spin glasses, whose equilibrium
thermodynamics are much more difficult to ascertain than those of
uniform ferromagnets, are (at least in some cases) {\it easier\/} to
analyze in the context of nature vs.~nurture.

\section{Persistence and Heritability in Low-Dimensional Uniform
  Ferromagnets}

The nature vs.~nurture question is intimately related to older notions of
{\it persistence\/}~\cite{DBG94}, defined as the fraction of spins that are
unchanged from their initial values at time $t$. This was found to decay as
a power law in a number of systems, in particular uniform ferromagnets and
Potts models in low dimensions, and the associated decay exponent
$\theta_p$ is known as the ``persistence exponent''.

In a similar manner, one can define a ``heritability
exponent''~\cite{YMNS13} as follows: prepare two Ising systems with
the same initial configuration but then allow them to evolve
independently using zero-temperature Glauber dynamics. The spin
overlap between these ``twin'' copies, with the same initial condition
but two different dynamical realizations, was found (after averaging
over many trials and different initial conditions) to decay as a power
law in time~\cite{YMNS13}. This spin overlap, which we refer to as the
``heritability'', is essentially the same as $q^t$. The exponent
$\theta_h$ associated with the power-law decay of heritability is the
``heritability exponent''.

Heritability defined in this way is in some sense the opposite of ``damage
spreading''~\cite{Creutz86,SSKH87,Grassberger95}; the latter involves
starting with two slightly different initial configurations and letting
them evolve with the {\it same\/} dynamical realization. The extent of the
spread of the initial difference throughout the system is then measured.

The persistence and heritability exponents can be computed exactly in
the $1D$ uniform Ising ferromagnet. It was shown in~\cite{DHP95,DHP96}
that $\theta_p=3/8$ for this system. On the other hand, it can be
shown that $\theta_h = 1/2$, as discussed in~\cite{YMNS13}, by using
the mapping to the voter model and coalescing random walks (see,
e.g.,~\cite{DHP96,FINS01}).

While the persistence and heritability exponents differ in one dimension,
they may be identical in the $2D$~uniform ferromagnet, where numerical
simulations yield $\theta_p=0.21\pm0.02$~\cite{Stauffer94,J99} and
$\theta_h=0.22\pm0.02$~\cite{YMNS13}. Whether the two exponents are exactly
the same, or simply close but not identical, remains to be understood.

\section{Positive temperature}

Does the preceding discussion have anything to say about what happens
at nonzero temperature?  Here one needs to study the behavior of
positive temperature Gibbs states and the local order parameter,
rather than that of single spin configurations.  Construction of the
appropriate dynamical measures, analysis of their evolution, and
relation to pure state structure are extensively discussed
in~\cite{NS99}.  Here we mention only a few relevant results.

The categorization into types ${\cal I}$, ${\cal F}$, and ${\cal M}$
is specifically tailored to zero temperature and needs to be modified
at positive temperature.  In the latter case, one can still define
local equilibration, in the sense that, on any finite lengthscale, the
system equilibrates into a {\it pure state\/} after a finite time
(depending on all of the usual culprits), in the sense that interfaces
cease to move across the region after that time.  If finite regions
exist without a corresponding finite equilibration timescale, then LNE
occurs.

A main result of~\cite{NS99} is relevant to spin glasses in
particular: if only a single pair, or countably many pairs (including
a countable infinity) of pure states exists (with fixed ${\cal J}$),
and these all have nonzero Edwards-Anderson~(EA) order
parameter~\cite{EA75}, then LNE occurs.  A corollary is that if LNE
does {\it not\/} occur, and the limiting pure states have {\it
  nonzero\/} EA order parameter, then there must exist an {\it
  uncountable\/} infinity of pure states, with almost every pair
having overlap zero.

One consequence of these results is that LNE occurs at positive
temperature (with $T<T_c$) in the $2D$ uniform ferromagnet and
(presumably) random Ising ferromagnets for $d<5$.  Because the number
and structure of pure states at positive temperature in Ising spin
glasses is unknown for $d\ge 3$ (and, from a rigorous point of view,
unproved even for $d=2$), occurrence of LNE there remains an open
question.

\section{Open problems}

The behavior of homogeneous and disordered Ising spin systems in one
and two dimensions is now relatively well understood. Beyond that,
however, most questions remain open. Do uniform ferromagnets belong to
class ${\cal F}$, ${\cal I}$, or ${\cal M}$ in dimension three and
higher? If ${\cal F}$, what is the value of $q_D$? If not, is weak LNE
or CTD displayed, and what is the value of the heritability exponent?

The relationships among heritability, persistence, and damage
spreading form an interesting set of open problems as well. Are the
heritability and persistence exponents the same in the $2D$
ferromagnet on a square lattice, and if so, why? What about higher
dimensions and other models? It would be interesting to study these
relations in two and higher dimensions and work out the connections
between these different but related quantities.

\subsubsection*{Acknowledgments.} The authors thank Jon~Machta,
Jing~Ye, Vladas~Sidoravicius, and P.M.C.~de~Oliveira for fruitful
collaborations on questions of nature vs.~nurture. This research was
supported in part by US~NSF~Grants~DMS-1207678 and~OISE-0730136.

\end{document}